# The Giving Game

*Dr. W.P. Weijland*
Delft University of Technology
May 2021

**Abstract.** Assume you are with N players in a room. One player has a ball. Once you have the ball you must choose a player – not being yourself – and pass the ball to her/him. The game repeats itself from then onwards. Every time you receive the ball you get a point. Question: which strategy is optimal to receive as many points possible?

One strategy would be to play the ball to the player from whom you received the ball most times in the past, assuming this creates the best chance to receive the ball in return. Receiving the ball creates an increase of preference for the submitting player who has just granted you a point. This seems rational to do.

A game described above is called a Giving Game. This paper reveals the fundamental structure of such a game and how the chosen strategy works out. Main result is that the game necessarily stabilizes into a repetitive pattern within a subgroup – in this case: a pair – of players. Furthermore, I will show that the path towards stabilization consists of a series of elementary cycles, each further enhancing the preferred pair.

Example applications are where computers are sharing processing power for complex calculations, or when commodity traders are making transactions in some professional community. The name 'Giving Game' has a positive connotation. However, it may equally well be viewed as a basic model of clientelism or corruption.

**INTRODUCTION**

Assume we play the following game. We have N agents and one token. Initially, the token is assigned to one of the agents: the initial agent. The game proceeds to the next step as follows: the agent with the token in possession submits its token to some other player, thus not being itself. Aim of the game is to maximize the number of tokens received over a certain period.

We call this game The Giving Game.

Once holding the token a strategy for any agent could be to assign the token to the agent from whom it has received the ball the most times in the past. The idea is to stimulate loyalty of those agents that are inclined to transfer the token to you. We will discuss how this strategy works out.

Firstly, imagine we play the Giving Game without any history: no one has received any token from anyone else in the past. The initial agent passes its token to some other agent. This agent now has a preference for the initial agent and will return the ball to it. After the token has returned to the initial agent both agents have a mutual preference for each other: both have received a token from the other in the past. So, the token will be passed back and forth between the two players endlessly. This instance of the Giving Game – without history – is called the *trivial game*.

The trivial game immediately alternates between two agents, one of which is the initial agent. We refer to such pair of agents as a *stability pair*. The term 'stability' refers to the fact that the path eventually repeats itself in a forced loop (in this case between two agents).

The question rises what happens when the agents *do* have a history. For each agent such history consists of a count of tokens received from each of the other agents. The remainder of this paper is devoted to proving that in such cases the game still stabilizes into a stability pair, be it after some steps (Stabilization Theorem II.5). We also analyze which pairs of agents are potential stability pairs and how such stability pair emerges from the initial steps in the game (Cycle Theorem VI.6).

**I       DEFINITION**

Let us further formalize the definition of a Giving Game, starting from N agents *A, B, C, ...* and one token. We assume each agent keeps track of a list of N-1 preference values (non-negative integer numbers) assigned to every other agent. The game proceeds to the next step as follows:

(1) the agent with the token in possession (the *submitting agent*) submits its token to some other agent (the *receiving agent*) with maximal preference value in the list;

(2) the receiving agent increments the preference value of the submitting agent in its list, and becomes submitting agent of the next step.

If we view the preference list of each agent as a column of values then these columns together form a *preference matrix*. Such matrix **M** may look as follows:

|   | *A* | *B* | *C* | *D* |
|---|---|---|---|---|
| *A* |   | 0 | 1 | 0 |
| *B* | 2 |   | 2 | 0 |
| *C* | 1 | 1 |   | 1 |
| *D* | 2 | 0 | 4 |   |

*Figure 1. A preference matrix.*

Submitting agents are in the top row. Receiving agents in the left column. Note that the diagonal cells have no preference values reflecting the fact that submitting agents cannot submit tokens to themselves.

NOTATION        Throughout this article we will write *(X, Y)* for the matrix cell at the intersection of column *X* and row *Y*.

As a helpful tool, at the start of any step in the game we color the submitting agent in red, cells with maximum column value in blue and all other cells white, such as in:

|   | A | B | C | D |
|---|---|---|---|---|
| A |   | 0 | 1 | 0 |
| B | 2 |   | 2 | 0 |
| C | 1 | 1 |   | 1 |
| D | 2 | 0 | 4 |   |

*Figure 2. A colored preference matrix.*

In this particular example submitting agent *C* has a maximum column value at row *D* – i.e. at cell *(C, D)*) – and will select agent *D* to submit the token to. The matrix is then updated by incrementing the value of cell *(D, C)* with 1, and *D* becomes the next submitting agent.

NOTATION   We write |*(X, Y)*| for the value assigned to a cell, though we will also use abbreviations like '*(X, Y)* = 2', '*(X, Y)* is incremented by 1', or '*(X, Y)* is maximal in the column of *X*' if that does not cause ambiguity. *(Y, X)* is referred to as the *twin cell* of *(X, Y)*[1]. |*(Y, X)*| is the *twin value* of *(X, Y)* and the color of *(Y, X)* *twin color* of *(X, Y)*.

Following the notions in the above, a giving game is defined by some initial pair $(A_0, \mathbf{M}_0)$ consisting of the initial agent and the initial preference matrix. From this initial pair the game develops into a *game sequence* $(A_i, \mathbf{M}_i)_{i \geq 0}$ where each next pair fits with the rules of the game. To be more precise:

DEFINITION I.1   Given a finite number of N agents, a *Giving Game* is defined by a pair $(A_0, \mathbf{M}_0)$, where $A_0$ is some agent – the *initial* agent – and $\mathbf{M}_0$ is some *preference matrix*: an NxN matrix with non-negative integer values assigned to each cell except for those on the diagonal.

DEFINITION I.2   Let $\mathbf{G}=(A_0, \mathbf{M}_0)$ be a Giving Game. A *game sequence* of **G** consists of an infinite sequence $(A_i, \mathbf{M}_i)_{i \geq 0}$ where all $A_i$ are agents and all $\mathbf{M}_i$ are preference matrices, and for all *i*:
a) $(A_i, \mathbf{M}_i)=\mathbf{G}$ for *i*=0.
b) |$(A_i, A_{i+1})$| is maximal in the column of $A_i$ in $\mathbf{M}_i$, i.e.: $(A_i, A_{i+1})$ is blue in $\mathbf{M}_i$.
c) |$(A_{i+1}, A_i)$| in $\mathbf{M}_{i+1}$ has incremented by 1 compared to its value in $\mathbf{M}_i$.
d) All other corresponding cell values in $\mathbf{M}_i$ and $\mathbf{M}_{i+1}$ are equal.

Agent $A_i$ is referred to as the *submitting agent* of step *i*, $A_{i+1}$ as its *receiving agent*, and cell $(A_i, A_{i+1})$ as *the selected cell* of step *i*. Note that in Definition I.2 the cell $(A_{i+1}, A_i)$ is the twin cell of the selected cell. In other words: the selection of a blue cell $(A_i, A_{i+1})$ at step *i* causes its twin cell to be incremented by 1.

A Giving Game **G** is *deterministic* if it has only one game sequence. Otherwise it is non-deterministic, implying that at some step *n*≥0 of some game sequence the submitting agent has a choice between two potential receiving agents (i.e. has at least two blue cells in its matrix column).

A handy notion related to a game sequence is that of a game path:

DEFINITION I.3   A (finite or infinite) sequence $(A_i)_{i \geq 0}$ is called a *game path* of **G** if all $A_i$ are subsequent submitting agents of some game sequence of **G**. If *A, B, C, D,…* are agents we may also write *ABCD…* as a shorthand for a (finite segment of) a game path.

---
[1] So the twin of a cell is equal to its diagonal mirror image in the matrix.

Clearly, not every sequence of agents is a viable game path since the initial matrix $\mathbf{M}_0$ (and thereby all subsequent preference matrices) puts restrictions on whether such sequence fits with Definition I.2.

## II  STABILITY

Let us go back to the trivial game in the introduction. It is easy to see that this is equivalent to the Giving Game with all values in the initial preference matrix put at zero.

In this section we will generalize this result to hold for any Giving Game.

NOTATION     Let $\mathbf{S} = (S_i)_{i \geq 0}$ be a sequence then for $n \geq 0$ we use the following abbreviations:
   a) $n|\mathbf{S} = (S_i)_{i \geq n}$, hence: $0|\mathbf{S} = \mathbf{S}$;
   b) $\mathbf{S}|n = (S_i)_{0 \leq i < n}$ and $\mathbf{S}|0 = \emptyset$, i.e.: the empty sequence.

So $\mathbf{S}|n$ is the segment of the first $n$ items in the sequence. $n|\mathbf{S}$ is the remaining infinite tail of the same sequence $\mathbf{S}$.

DEFINITION II.1     A pair of agents $\{A, B\}$ is a *stability pair of game* $\mathbf{G}$ if for some game path $\mathbf{S}$ of $\mathbf{G}$ there exists $n$ such that either $n|\mathbf{S} = ABABAB...$ or $n|\mathbf{S} = BABABA...$ Furthermore, a pair of agents $\{A, B\}$ is a *stability pair of a preference matrix* $\mathbf{M}$ if it is a stability pair of $\mathbf{G}=(C, \mathbf{M})$ for some agent $C$.

LEMMA II.2     Let $\mathbf{G}=(A, \mathbf{M})$ be a Giving Game. Suppose $(A, B)$ and $(B, A)$ are both blue cells in $\mathbf{M}$. If at the initial step $A$ selects $B$ then the game path is equal to $ABABAB...$

PROOF  Since $(A, B)$ is blue, $B$ is maximal in the column of $A$ and according to Definition I.2b agent $A$ may choose $B$ at step 0. Following Definition I.2c the value of $(B, A)$ will be incremented. Since it is blue in $\mathbf{M}$ it will become *single maximum* in the column of $B$. But then in the next step $B$ must choose $A$ thereby increasing the value of $(A, B)$ which in turn becomes single maximum in the column of $A$. By induction the game path is equal to $ABABAB...$ □

Note that this Lemma is a generalization of the argument for the trivial game at the introduction.

LEMMA II.3   Let $\mathbf{G}$ be a Giving Game. A pair of agents $\{A,B\}$ is a stability pair of $\mathbf{G}$ if and only if there exists a game path of $\mathbf{G}$ containing $ABA$.

The proof resembles the proof of Lemma II.2 and is left to the reader. Later we will see that the Lemma can be further generalized to "...containing both steps $AB$ and $BA$". Also this proof is left to the reader.

LEMMA II.4     Let $\mathbf{G}=(A, \mathbf{M})$ be a Giving Game. If $(A,B)$ and $(B,A)$ are both white cells in $\mathbf{M}$ then $ABA$ is *not* a segment of a game path of $\mathbf{G}$.

PROOF  It follows from Definition I.2 that a cell value can only be incremented if its twin cell is blue. Thus the values of $(A,B)$ and $(B,A)$ remain unchanged, and since cell values cannot decrease none of these cells can ever turn blue. □

THEOREM II.5 (STABILIZATION) Every game path eventually alternates between two agents.

PROOF  At every step in the game path a blue cell is selected in some column of a game matrix. If its twin cell is also blue, then Lemma II.2 applies. So assume at all steps twin cells are white. Let MAX be the maximum of all values in the initial matrix $\mathbf{M}$. In each step a white cell is incremented and since the value of that white cell is smaller than MAX, the maximum value in any subsequent matrix is also equal to MAX. Now if we have N agents, we have N.(N-1) cells in the matrix. So after N.(N-1).MAX+1 increments, at least one of the twin cells must contain a value ≥MAX. But then it cannot be white. Contradiction. □

### III  COLORED PAIRS

From the Stabilization Theorem II.5 we know that every game path stabilizes eventually. We may wonder how this stability pair is ultimately created in the game. We refer to the long alternating tail of a game path as the *stability* phase of that path. The initial part of a game path until the stability phase is referred to as the *erratic phase* emphasizing its seemingly hectic nature. It is where choices of individual agents in the game determine which stability pair eventually emerges.

Before we formally define the stability and erratic phase, let us take a closer look at how the Giving Game is played.

EXAMPLE  Let us look again at the matrix in Figure 2:

|   | A | B | C | D |
|---|---|---|---|---|
| A |   | 0 | 1 | 0 |
| B | 2 |   | 2 | 0 |
| C | 1 | 1 |   | 1 |
| D | 2 | 0 | 4 |   |

- Starting from *C* the game immediately stabilizes into *CDCDCD...* since *C* has only one maximum to choose from (i.e.: at row *D*), so we get *CD...* . Since both *(C,D)* and *(D,C)* are blue cells Lemma II.2 applies.
  A twin pair of blue cells is called a *blue pair*. Lemma II.2 states that if the submitting agent selects a cell from a blue pair, then the game stabilizes immediately.

- Starting from *B* the game proceeds as *BCDCDCD...* . Note that while *B* selects *C*, the cell value of *(C,B)* is incremented with 1 (increments from 2 to 3). However, since the maximum value of *C*'s column is 4, the color of *(C,B)* remains white.
  A twin pair of which one cell is blue and the other is white is called a *turquoise pair*. Of such pair only the blue cell can be selected by the submitting agent (by definition) and its twin cell may stay white (as in this case) or turn blue (for example if the maximum value in *C*'s column would have been 3), thus creating a blue pair in the next step's preference matrix.

- Starting from *A* things are getting slightly more complex. The game proceeds in either of two ways: *AB...* or *AD...* . Both cells *(A,B)* and *(A,D)* are blue and *(B,A)* and *(D,A)* are white, so in either case *A* selects a turquoise pair.
  The two cases develop as follows:
  1. In the case of *AB...* note that *(B,A)* is incremented with 1 and turns blue. Now *B* chooses either *A* (with which it then forms a blue pair) to stabilize at *ABABAB...* (stability pair {*A,B*}), or *C* (with which it forms a turquoise pair) to stabilize at *ABCDCDCD...* (stability pair {*C,D*}).

2. In the case of *AD…* we find that the value of *(D,A)* increments from 0 to 1 thereby turning blue. Now *D* has a choice between *A* and *C* both of which form a blue pair with *D*. So the game stabilizes either as *ADADAD…* or as *ADCDCDC…* .

We count four stability pairs of this game out of six possible combinations of two agents. Note that {*A,C*} and {*B,D*} are the pairs missing: they are twin pairs of white cells, referred to as *white pairs*, and by Lemma II.4 they are no stability pairs.

[END EXAMPLE]

From the Lemmas II.2 and II.3 three we learn that blue pairs are stability pairs whereas white pairs are not. The kind missing are the turquoise pairs. We will use this observation in a few moments.

## IV    FRAMES

Before we continue let us present a useful simplification of the Giving Game that helps us find stability pairs. We introduce the concept of a *frame* as a preference matrix without values. Such matrix inherits the structure of the preference matrix, with the same agents and the same cell colors but with no cell values to increase. It is referred to as the *frame* of a preference matrix and looks as follows:

|   | A | B | C | D |
|---|---|---|---|---|
| A |   |   |   |   |
| B |   |   |   |   |
| C |   |   |   |   |
| D |   |   |   |   |

*Figure 3. The frame of the preference matrix in Figure 2.*

We can generalize the concept of a game path to that of a *frame path* as follows:

DEFINITION IV.1    Let **F** be the frame of some preference matrix. A *frame path of* **F** consists of an infinite sequence $(A_i)_{i \geq 0}$ where all $A_i$ are agents in **F**, and $(A_i, A_{i+1})$ is blue in the column of $A_i$ in **F**. A *finite frame path* is a finite segment of a frame path.

Note that the frame **F** remains unchanged along its frame paths. For that reason the concatenation of two (finite) frame paths is again a frame path.

DEFINITION IV.2    Let **F** be a frame. *B is reachable from A* – notation: $A \sqsupseteq_F B$ – if there exists some finite frame path *A…B* of **F**.

DEFINITION IV.3    *A* and *B* are *equivalent* – notation: $A \equiv_F B$ – if $A \sqsupseteq_F B$ and $B \sqsupseteq_F A$.

LEMMA IV.4    $\equiv_F$ is an equivalence relation.

The proof is left to the reader. The equivalence relation $\equiv_F$ creates equivalence classes ***A, B, C***, … (sets of equivalent agents) that are partially ordered as follows:

DEFINITION IV.5    Let **F** be a frame and ***A*** and ***B*** sets of equivalent agents in **F**, then we define: $\boldsymbol{A} \sqsupseteq_F \boldsymbol{B}$ if for some agents $A \in \boldsymbol{A}, B \in \boldsymbol{B}$: $A \sqsupseteq_F B$.

For each agent $A \in \boldsymbol{A}$ we thus obtain a tree structure with ***A*** at the top node and classes of reachable agents at lower nodes. Along each path in the tree classes are mutually exclusive. Edges in the tree are one direction only: lower nodes are reachable from the higher but not the other way around. For example, starting from *A* the structure of the frame of Figure 3 looks like:

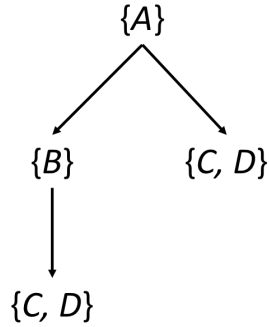

*Figure 4. Equivalence classes in Figure 3 in a tree.*

Note that if we would put {*C,D*} at the top of the tree we would just have one single node, since no other classes are reachable from {*C,D*}. Instead of a tree, we can also use the structure of a *directed graph* as in Figure 5

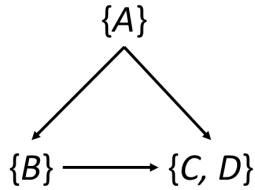

*Figure 5. Equivalence classes in Figure 3 in a directed graph.*

Such directed graph consists of all equivalence classes of **F**, not only those reachable from the top.

DEFINITION IV.6    A finite game (frame) path is *elementary* if all its agents occur only once.

LEMMA IV.7    Let **F** be a frame. If there exists a frame path from *A* to *B* (i.e.: $A \sqsupseteq_\mathbf{F} B$) then there also exists an *elementary* frame path from *A* to *B*.

PROOF We provide an intuitive proof. Consider a frame path *A…B*. Suppose this frame path has a multiple occurrence of some agent *C*, then we have: *A…C…C…B*. Since **F** remains unchanged along the path, at each step each agent has the same choice options to continue. But then we can shortcut the path to obtain *A…C…B*, reducing *C…C* to a single occurrence *C*. By repeating this argument for every multiple occurrence of some agent we find an elementary path *A…B*. □

The segment *C…C* that is eliminated from the path is referred to as a *cycle*. Cycles play an important role in the creation of new stability pairs as we will see later.

DEFINITION IV.8    Let **F** be a frame. A *cycle* is a finite frame path from some agent to itself.

It follows directly from Lemma IV.7 that each cycle can be reduced to an elementary cycle starting from the same agent.

## V  ERRATIC PHASE AND STABILITY PHASE

In section III we already spoke of the concept of erratic and stability phase. We formalize these notions in a definition.

DEFINITION V.1    Let **G**=(*A*,**M**) be a Giving Game and **S**=(*A_i*, **M**_i)_{i≥0} be a game sequence of **G**. Let $n≥0$ be the smallest $n$ such that *(A_{n+1}, A_n)* is blue in **M**_n, then the subsequence **S**|n is the *erratic phase* of **S** and n|**S** is the *stability phase* of **S**.

By Theorem II.5 that such $n$ exists for every game sequence **S**, so the definition is sound. It says that the stability phase starts at the first step $n$ in **S** with a blue twin. It follows from Lemma II.2 that from this step $n$ all pairs selected thereafter are blue since all cells $(A_n, A_{n+1})$ are blue (by Definition I.2). As a consequence we have:

COROLLARY V.2    In the stability phase all selected pairs are blue and in the erratic phase all selected pairs are turquoise (see Section III).

In the erratic phase every step consists of selecting a blue cell from the column of the submitting agent such that its twin cell is white.

LEMMA V.3    Let **S**=(*A_i*, **M**_i)_{i≥0} be a game sequence. If for some $n$ cell $(A_n, A_{n+1})$ is blue in **M**_n and white in **M**_0 then step $n$ is in the stability phase of **S**.

PROOF  Let *(B, C)* be blue in **M**_n and white in **M**_0. Then for some $k<n$ cell *(B, C)* must turn blue for the first time: it is white in **M**_k and blue in **M**_{k+1}. In that case the value of cell *(B, C)* must have incremented at step $k$, which only results from *C* selecting *B*. So *(C, B)* must be blue in **M**_k (in order to select *B*) and it will stay blue in **M**_{k+1}, since *(B, C)* and *(C, B)* cannot change color at the same step. Now both *(B, C)* and *(C, B)* are blue in **M**_{k+1}. By Definition V.1 step $n$ is in the stability phase. □

This lemma implies that during a game sequence blue cells from the initial matrix cannot change color until the sequence has reached its stability phase. This leads to the following insight:

LEMMA V.4    The erratic phase of a game path of **M** is also a frame path of **M**.

PROOF  By Lemma V.3 we know that no blue cells from the initial matrix **M** turn white during the erratic phase. All newly created blue cells originate from white twin cells that are incremented at some step. At the step they are created they form a blue pair with the (blue) selected cell. By Definition V.1. this step cannot lie in the erratic phase. □

In summary we look at the following. At each step of the game the submitting agent selects a receiving agent by picking a blue cell from its column while updating its twin cell (i.e.: adding 1 to its value) before the next step begins. If the twin of this blue cell is also blue, then we are in the stability phase (Definition V.1). As long as it is white we are in the erratic phase. While the game progresses in the erratic phase, white cells may turn blue but never the other way around: blue cells cannot turn white, unless in the stability phase (Lemma V.3).

## VI  CYCLES

Suppose we put ourselves the task to find out whether {*A, B*} is a stability pair, given some initial matrix **M**. How would we be able to decide? Some basic cases are trivial. For instance, if {*A, B*} is a blue pair in **M** then it is a stability pair starting from initial agent *A* or *B*. It is also a stability pair starting from some other initial agent *C* provided *A* or *B* is reachable from *C* (Definition IV.2). If {*A, B*} is a white pair, we know that it *cannot* be a stability pair regardless the initial agent (Lemma II.4). So the only case left is when {*A*, *B*} is a turquoise pair in **M**.

(\*)     In this section let us assume *(A, B)* is a blue cell in **M** and twin cell *(B, A)* is white.

We are looking for game paths stabilizing at *ABABAB…* or *BABABA…* Note that any such game path must have some step where *(B, A)* turns blue (since otherwise step *BA* would not be valid). This implies that the value in cell *(B, A)* has 'caught up with' the maximum of column *B* during the erratic part of the game path. Values of blue cells can only change in the stability phase since they are the twin of some other blue cell and blue pairs only occur in the stability phase (Corollary V.2). So the maximum value that *(B, A)* catches up with is also a maximum of column *B* in **M**.

DEFINITION VI.1     The *gap* of a cell *(X, Y)* in a game matrix **M** – notation: $gap_\mathbf{M}(X, Y)$ – is equal to the difference between the maximum value in column *X* of **M** and the cell value of *(X, Y)*.

The gap of cell *(B, A)* can be closed (i.e.: become zero) during the game by selecting *(A, B)* sufficiently many times. In fact, since at every selection of *(A, B)* the gap decreases with 1 we need to select cell *(A, B)* precisely $gap_\mathbf{M}(B, A)$ many times in order make cell *(B, A)* turn blue. This leads to a simplification of the original question:

COROLLARY VI.2     A turquoise pair is a stability pair of a game (*C*, **M**) if and only if there exists a game path of (*C*, **M**) with at least $gap_\mathbf{M}(B, A)$ many occurrences of step *AB*.

EXAMPLE         Let us look again at the matrix in Figure 2:

|   | *A* | *B* | *C* | *D* |
|---|---|---|---|---|
| *A* |   | 0 | 1 | 0 |
| *B* | 2 |   | 2 | 0 |
| *C* | 1 | 1 |   | 1 |
| *D* | 2 | 0 | 4 |   |

Pair {*A*, *B*} is a turquoise pair with *(A, B)* blue and *(B, A)* white. Cell *(B, A)* has value 0. The maximum in column *B* is 1 at cell *(B, C)*. So the gap of *(B, A)* = 1 – 0 = 1. So we are looking for game paths with at least one occurrence of step *AB*. Obviously, in (*A*, **M**) we find such path: *AB* itself. However, in the games (*B*, **M**), (*C*, **M**) and (*D*, **M**) such path does not exist.
[END EXAMPLE]

So let us zoom in further on the structure of a game path with stability pair {*A*, *B*}. Such a path starts with an initial agent – say: *C* – and has at least $gap_\mathbf{M}(B, A)$ many occurrences of *AB*. So it looks like: *C…AB…AB…AB…ABABAB…* In the following we will define a reduction system to reduce any game path with stability pair {*A*, *B*} to a 'shorter version' of the same path. Such reduction makes use of reduction rules as follows.

REDUCTION SYSTEM VI.3     Let **M** be an initial matrix. Consider a game path of **M** with arbitrary initial agent and with stability pair {*A*, *B*}. We look at the following two-rule reduction system:

(1) If the initial segment of the path until the first occurrence of step *AB* is not empty, then we cut it off making the reduced game path start with *AB* while leaving the rest of the path unchanged.

(2) If between any two consecutive steps *AB...AB* some agent (including *A* or *B*) has a multiple occurrence of *C*, as in: *AB...CDEFC...AB*, then delete *CDEF* to produce the reduced form *AB...C...AB*.

[END REDUCTION SYSTEM]

Obviously, reduction rule (1) can only apply once to a game path. The second, however, can be applied as many times until all multiple occurrences of agents between consecutive steps *AB* have been eliminated.

LEMMA VI.4        Reduction system VI.3 preserves the following properties:
(i)     the reduced game path is again a valid game path;
(ii)    the reduced game path maintains stability pair {*A, B*}.

PROOF  Let **S** be a game path of (*C*, **M**) with stability pair {*A, B*}. Suppose **S'** is the erratic phase of **S** and <u>**S**</u> its stability phase. Hence: **S** = **S'**<u>**S**</u>. By Lemma V.4, **S'** is also a frame path of **M**. The proof consists of five claims:

CLAIM 1: Reduction system VI.3 only applies to segments of **S'**.
Rule (i): If any step before the first occurrence of *AB* has a blue twin, it is in the stability phase of **S** (Corollary V.2). Then it must be either *AB* or *BA*. Step *AB* cannot occur before its first occurrence. Since *(B, A)* is white in **M** (by (*)) step *BA* cannot be a valid step before the first occurrence of *AB*. Contradiction.
Rule (ii): The stability phase has only occurrences of *AB* with no other agents between them. So rule (2) cannot apply outside the erratic phase of **S**.

So let us assume Reduction system VI.3 reduces **S'** to **S''**.

CLAIM 2: **S''** is a turquoise frame path of **M**.
Rule (i): Trivial.
Rule (ii): If we cut out a cycle segment from **S'** the resulting path **S''** only consists of steps from **S'**. So **S''** remains a turquoise frame path of **M** [2].

CLAIM 3: **S''** is a turquoise game path of **M**.
Since **S''** is a turquoise frame path it only selects blue cells in **M** with white twin cells. Suppose that after some steps in **S''** one of these twin cells – say: *(Y, X)* – has turned blue. Then blue cell *(X, Y)* must have been selected at least $gap_M(Y, X)$ many times. The number of occurrences of *(X, Y)* in **S''** cannot be higher than that in **S'** so *(Y, X)* must also turn blue at some step to form a blue pair in **S'**, contradicting that **S'** is erratic.

CLAIM 4: The last agent of **S''** is equal to that of **S'**, or **S'** is fully deleted.
Rule (i): Trivial.
Rule (ii): Any cycle *C...C* deleted from the erratic phase is replaced by *C* and therefore the last agent of the erratic phase remains in place.

CLAIM 5: **S''**<u>**S**</u> is a game path of **M** with stability pair {*A, B*}.
**S** has stability pair {*A, B*} so its erratic phase **S'** contains precisely $gap_M(B, A)$ many occurrences of step *AB* (Corollary VI.2). Since these occurrences are preserved by Reduction system VI.3 we find equally many occurrences of *AB* in **S''**. Therefore {*A, B*} is a blue pair after the last step of **S''**. The last agent of **S'** – say *C* – is unaffected by Reduction system VI.3. Since it is in the erratic path **S''** no blue cell in *C's* column has change color, so *C* may choose *A* or *B*, whichever is first in the stability phase. So the concatenation **S''**<u>**S**</u> is a game path of **M** with stability pair {*A, B*}. □

---
[2] Note: if we would cut out non-cyclic segments from **S'** this may no longer hold, since we then may create steps in **S''** that are not present in **S'**.

DEFINITION VI.5    A game path of **M** with stability pair {A, B} is in *normal form* if none of the reduction rules in VI.3 apply.

The following Corollary follows immediately from the construction of Reduction system VI.3:

COROLLARY VI.6 (CYCLE THEOREM)
1. {A, B} is a stability pair of **M** if and only if some game path in normal form stabilizes at {A, B}.
2. A normal form path starts with its stability pair as the first step.
3. The erratic phase of a normal form is built from the concatenation of a series of elementary cycles AB…A.

All of the elementary cycles from Corollary VI.6.3 are present in the frame of **M**, so we do not have to look further than the initial matrix to identify them. Also note that agents in a cycle are equivalent (Definition IV.4) and so a normal form game path has only agents from one single equivalence class, as they all contain A and B.

It is not difficult to see that if we would change the order of occurrences of the elementary cycles in Corollary VI.6.3, then the sequence remains a valid game path stabilizing at {A, B}. So in the process of finding a game path stabilizing at {A, B} we need to focus on which cycles are involved and how many times each of them occurs (the 'power' of the cycle).

EXAMPLE    A Giving Game 'produces' new stability pairs (i.e.: not blue in the initial matrix) by running cycles. In Figure 6a below we illustrate how this works.

|   | A | B | C | D |
|---|---|---|---|---|
| A |   | 2 | 1 | 9 |
| B | 4 |   | 0 | 2 |
| C | 0 | 7 |   | 1 |
| D | 1 | 5 | 5 |   |

*Figure 6a: All agents in one cycle.*

Note that the frame of this matrix is fully deterministic: it does not leave choices open to agents. Also note that the matrix has no blue pairs. So any stability pair must be created throughout the game.

Further, check that *ABCDA…* is a cycle. Following the steps in this cycle the twins of the selected cells are incremented by Definition II.2. These incremented values are marked red in the left-hand matrix in Figure 6b:

|   | A | B | C | D |
|---|---|---|---|---|
| A |   | 3 | 1 | 9 |
| B | 4 |   | 1 | 2 |
| C | 0 | 7 |   | 2 |
| D | 2 | 5 | 5 |   |

|   | A | B | C | D |
|---|---|---|---|---|
| A |   | 4 | 1 | 9 |
| B | 4 |   | 2 | 2 |
| C | 0 | 7 |   | 3 |
| D | 3 | 5 | 5 |   |

|   | A | B | C | D |
|---|---|---|---|---|
| A |   | 5 | 1 | 9 |
| B | 4 |   | 3 | 2 |
| C | 0 | 7 |   | 4 |
| D | 4 | 5 | 5 |   |

|   | A | B | C | D |
|---|---|---|---|---|
| A |   | 6 | 1 | 9 |
| B | 4 |   | 4 | 2 |
| C | 0 | 7 |   | 5 |
| D | 5 | 5 | 5 |   |

*Figure 6b: All agents in one cycle (cont'd).*

Still, blue cells remain blue and white cells remain white. In the second run of the same cycle we obtain *ABCDABCDA…* – in shorthand: $(ABCD)^2 A…$ – and we arrive at the second matrix where the red twin cells are incremented again. At the third run we have $(ABCD)^3 A…$ and we obtain the third matrix with cell *(A, D)* turned blue reflecting the fact that its value now matches the maximum value in its column. Now, *A* has two options to choose from: either *AB…* or *AD…* . The first option *AB…* leads to a fourth run of the same cycle, so we have $(ABCD)^4 A…$ to arrive at the right-hand matrix. This time *A* only has one option to continue by choosing *D.* So the path stabilizes at *…ADADAD…* to form $(ABCD)^4 ADADADA…$ . It is easy to check that the second option *AD…* leads to the same stability pair to form $(ABCD)^3 ADADADA…$ .
[END EXAMPLE]

So, to decide whether *{A, B}* is a stability pair of a matrix **M**, we – theoretically – may wish to list all elementary cycles in **M** and see if we can add powers to them such that:
    (1) *{A, B}* occurs precisely $gap_\mathbf{M}(B, A)$ many times;
    (2) No pair *(X, Y)* occurs more than $gap_\mathbf{M}(Y, X)$ many times.
Let us take a closer look at the matrix in Figure 6a. The matrix has one elementary cycle. The steps in the cycle are *AB, BC, CD, DA* and the values in their twin cells *(B, A), (C, B), (D, C)* and *(A, D)* are 2, 0, 1, 1 respectively. The maximum values in their respective columns *B, C, D* and *A* are: 7, 5, 9 and 4. So the respective gaps (Definition VI.1) of the twin cells ('twin gaps') are 5, 5, 8, 3: the lowest being 3. This explains that after three runs of the cycle the first twin cell to turn blue is *(D, A)*.

This little example also shows that – in order to make *{A, B}* a blue pair – we cannot simply run the cycle 5 times: if we do, at the fourth run *{A, D}* would irreversibly deviate into stability and the game will never get back to *B*. We cannot run any cycle more times than its minimal gap of twin cells, unless it is *{A, B}* itself that has the minimal *twin gap*. In such case it is at *pole position* in the cycle. The minimal twin gap is the *order* of the cycle.

Lemma VI.7          If *{A, B}* is at pole position in some elementary cycle of **M**, then it is a stability pair.

If *{A, B}* is not in pole position at any elementary cycle then to determine whether it is a stability pair gets more tricky. One could think that we simply had to run each elementary pair *AB…A* as often as its order. Since all cycles contain *AB* we only need to check whether the sum of all those orders is at least $gap_\mathbf{M}(B, A)$. This would work indeed if all cycles would be disjoint in steps (apart from step *AB*, ofcourse). If not, running cycle 1 may lower the gap of steps in cycle 2 and thereby its order. In other words: orders of cycles are not mutually independent unless the cycles are disjoint. The tricky thing now is that we must find out which cycle to run how many times so as to maximize the overall number of runs.

I have not found an efficient algorithm to achieve this. I leave the question open here whether an algorithm exists with less than exponential complexity in relation to N.

Part of the complexity of the algorithm comes from the number of elementary cycles in a matrix **M**. This determines how many elementary cycles in a game we must evaluate. We have bad news:

Lemma VI.8          The number of elementary cycles may grow exponentially with *N*.

PROOF Consider the matrix in Figure 7. Let us say **M6** is the smaller matrix with six agents *A…F*. If we add column and row *G*, then we create a larger matrix **M7**, and also **M8** by adding *H*. This way we build a series of matrices **M***k* for any natural number *k*. All matrices only have turquoise pairs. Every next matrix inherits all cycles from the previous one. CLAIM: For all even *k* and every elementary cycle *AB…A* in **M***k* there is a unique elementary cycle in **M***k*+**1** that does not exists in **M***k*. Thus we prove that **M***k*+**1** has at least twice as many elementary cycles as **M***k*.

*Figure 7: Building matrices with only turquoise pairs.*

We will prove the CLAIM for *k*=6 and leave the general proof by induction to the reader. So we will prove that **M7** (with agents *A,…,G*) has at least twice as many elementary cycles as **M6**. Now let us say we have *odd* agents *A, C, E, G,…* and *even* agents *B, D, F, H…* Note that:
  I. Agent *B* can only select odd agents.
  II. In **M7** agent *G* can select all odd agents.

Assume we have an elementary cycle *AB…A* in **M6**. The cycle cannot be *ABA* since all pairs in **M6** are turquoise. Let *X* be such that the cycle in **M6** is equal to *ABX…A*. Clearly, *ABX…A* is also a cycle in **M7**. *X* is an odd agent, since it is selected by *B* (property I). In **M7** agent *G* can select *X*, since *X* is odd (property II). As *G* is odd, *B* can select *G* (property I). So *ABGX…A* is an elementary cycle in **M7**.

So each elementary cycle *AB…A* in **M6** has two unique images in **M7**: *AB…A* and *ABG…A*. But then **M7** has at least twice as many elementary cycles *AB…A* as **M6**. □

So simply listing all elementary cycles and find the optimal combination of powers is not the way to go for large N unless the frame of the matrix in which agents are ordered has some regular pattern to it.

**VII    AFTERMATH**

Back in 1980 my friend Jan Wielemaker – fellow member of the student society SSRE – revealed to me how social relations really work: 'every interaction with someone else leads to a social credit/debet that adds to a net positive or negative balance of favours'. The fact that I still remember his words means they made an impression, which – however – did not materialize in action until 2014, when the original idea of the Giving Game struck me as an rudimentary model of the phenomenon he pointed out.

At the time, I was working on a mathematical model to restore some equilibrium result between supply and demand in the context of a 'gift economy' ([4]). An old publication by Kiyotaki & Wright [6] inspired me to look at game theoretical models of exchange of goods. Thus the trivial model (section I) emerged as a rudimentary form of a giving game that provides stabilization, a property comparable to a supply and demand equilibrium. Non-trivial extension of this model provided the subject of this publication.

The concept of 'giving' in the sense of asynchronous transactions – comparable to promises [1], altruistic forms of giving at the benefit of a 'warm glow' [2, 3], or the 'favours' of Jan Wielemaker – can be posed opposite to a 'tit for tat' exchange trade economy where every transaction consists of an economically neutral transfer of assets. The name 'Giving Game' has a positive connotation. However, it may equally well be viewed as a basic model to describe the mechanics of clientelism or corruption.

The Giving Game as an attempt to understand the notion of 'giving' in the economical sense. It allows for many topics of future research. Let me briefly mention some:

*Stability and Community Effect*
The standard game presented in this paper has the pivotal stabilization theorem (Theorem II.5) as a center characteristic. The resulting stability pair may be looked at as a 'community' that eventually holding the token for itself. My vision is that this concept can be extended to more complex games while still preserving the existence of communities as objects of stabilization. Such complex games may have added features, such as for example:
- Multiple tokens, possibly each with a different 'weight' reflected by different increments of cell values;
- Introduction of non-durable tokens in this game, for example by adding a decay period of some fixed number of steps in the game;
- A 'hold' option for submitting or receiving agents: they may be temporarily unavailable for receiving (saturation) or submitting (keep stock) tokens.

One aim of my research is to identify under which conditions we can restore the Community Effect in the form of a community of agents circulating certain tokens amongst themselves (a stability pair is a simplified version of such community).

*Economic modelling*
The game extensions in the previous paragraph already create a micro-economic model of value transactions. A way to look at the preference matrix is like that of a historical balance of transactions that have taken place in the past (see Figure 6). Column values represent *tokens received / consumed* whereas row values represent *tokens submitted / produced*. So, for each agent rows represent the distribution of production units across all agents. Columns represent the distribution of units consumed across all agents.

|   | A | B | C | D | Total Production |
|---|---|---|---|---|---|
| A |   | 0 | 1 | 0 | **1** |
| B | 2 |   | 2 | 0 | **4** |
| C | 1 | 1 |   | 1 | **3** |
| D | 0 | 0 | 4 |   | **4** |
| Total Consumption | **3** | **1** | **7** | **1** | **12** |

*Figure 6: Preference matrix as a scheme of production and consumption.*

In Figure 6 total production and consumption are in balance, which in general may not be the case.

*Game Theory*
In The Giving Game we have introduced a decision rule for submitting agents to select a receiving agent. In what sense can this rule be regarded as an 'optimal strategy' in the game theoretical sense of the word? Can we define a game space of potential strategies where such decision rule forms a Nash Equilibrium, i.e.: a strategy that no individual agent can improve?

*Computational Complexity*
I have shown that any game path leading to a stability pair in the game has a normal form (Corollary VI.6). Can we establish some efficient (polynomial) algorithm to calculate that such path exists, given a matrix **M** and a pair {*A, B*}? Or can we prove it equivalent to some well-know problem with exponential complexity and prove the opposite?

*Crypto Currency*
A preference matrix is, in fact, a simplified accumulated version of a ledger of all past transactions [7, 8]. See also Figure 6. Such ledger is a common element of many – if not all – crypto currencies. Can we relate blockchain technology to the Giving Game described here?

*Applications*
I imagine several applications of the theory in this paper. I mention two:
- Firstly, in a network of distributed computing systems individual processors are at continuous search of idle processing power elsewhere in the network to commission partial tasks of computation to them [5]. A strategy of mutual preference – based on historical exchange of processing power – may enhance their chances to succeed their tasks in the future.
- Secondly, think of professional commodity traders trading vast amounts of stock with other traders every day. Their individual mission is to buy at the lowest price and sell at the highest. However, at a higher level patterns of preference may emerge between traders that share a track record of positive mutual benefits (e.g.: "*he saved my ass the last few times*") causing a preference bias towards each other at a lower than 'best price' level. This may cause inefficiency in a – relatively small – community of traders worldwide.